\documentclass[12pt, a4paper]{article}
\usepackage{a4wide}
\usepackage[normal]{subfigure}
\usepackage{graphicx}

\renewcommand{\thesubfigure}{\thefigure\alph{subfigure}}
\makeatletter
\renewcommand{\@thesubfigure}{\thesubfigure:\space}
\renewcommand{\p@subfigure}{}
\makeatother

\begin{document}

\title{Global quantum Hall phase diagram\\ from visibility diagrams}
\author{F. Chandelier$^a$, Y. Georgelin$^a$, T. Masson$^b$, J.-C. Wallet$^a$}

\maketitle
\begin{center}
$^a$ Groupe de Physique Th\'eorique,\\
Institut de Physique Nucl\'eaire\\
F-91406 Orsay Cedex, France\\
\bigskip
$^b$ Laboratoire de Physique Th\'eorique (UMR 8627)\\
B\^at 210, Universit\'e Paris-Sud Orsay\\
F-91405 Orsay Cedex
\end{center}

\bigskip
\begin{abstract}
We propose a construction of a global phase diagram for the quantum Hall effect. This global phase diagram is based on our previous constructions of visibility diagrams in the context of the Quantum Hall Effect. The topology of the phase diagram we obtain is in good agreement with experimental observations (when the spin effect can be neglected). This phase diagram does not show floating.
\end{abstract}

\vfill
LPT-Orsay 01-64

\newpage

\section{Introduction}

In previous papers \cite{GW1, GMW1, GMW2, GMW3}, we succeeded to classify the quantum Hall states by some visibility diagrams, and confronted them to some experimental features of the Quantum Hall Effect (QHE) \cite{KDP, TSG, PG}. We were able to reproduce the Jain and Haldane hierarchies \cite{J, H}, and all the other observed hierarchical structures of the Hall states. A plot of the Hall resistivity versus the magnetic field was derived from these diagrams, which was in good agreement with the experimental Hall resistivity plot \cite{GMW1}. Then, the stripe structure of these diagram was compared to experimental measurements of the mesoscopic conductance fluctuations in silicon MOSFETS \cite{GMW3}. All these previous theoretical and experimental considerations gives us now a clearer physical interpretation of the proposed visibility diagrams. In particular, we were able to define a mapping between the axis of these diagrams and physical quantities as it will be recalled below.

In this letter, we will assume that this mapping is relevant. Then, it can be used to construct a global phase diagram for the QHE. This global phase diagram gives a hint of the respective positions of the integer and fractionnal states. It is clearly in favor of the ``non floating'' scheme advocated by some authors \cite{HSSTX, KMFP, WXNJ}, and is in good agreement with many experimental observations made so far, at least for monolayer samples and if the effects of spin is neglected.

\section{The visibility diagrams}

We would like to recall here the general features of the visibility diagram, and their physical interpretation as has been analysed in \cite{GW1, GMW1, GMW2, GMW3}. The visibility diagrams were constructed originally from considerations about some modular symmetries in the QHE \cite{GW1, GMW1}, and specifically the possible relevance of the modular group $\Gamma(2)$ as a classifying group for the quantum Hall states. These diagrams can in fact be completely characterized independently of any modular group. Because modular symmetry considerations will not be essential in the following analysis, we choose to introduce them without any reference to these groups.

From a mathematical viewpoint, the construction of the visibility diagrams is easy to achieve and is based on an famous arithmetical theorem which states that if $p$ and $q$ are relatively prime integers, then there exists two (necessary relatively prime) integers $a$ and $b$ such that $pa - qb = 1$, and two others integers $a'$ and $b'$ (also relatively prime) such that $pa' - qb' = -1$. 

\begin{figure}[ht]
\centering
\subfigure[The odd-odd visibility diagram.]{
\label{fig1}
\includegraphics[width=0.47 \textwidth]{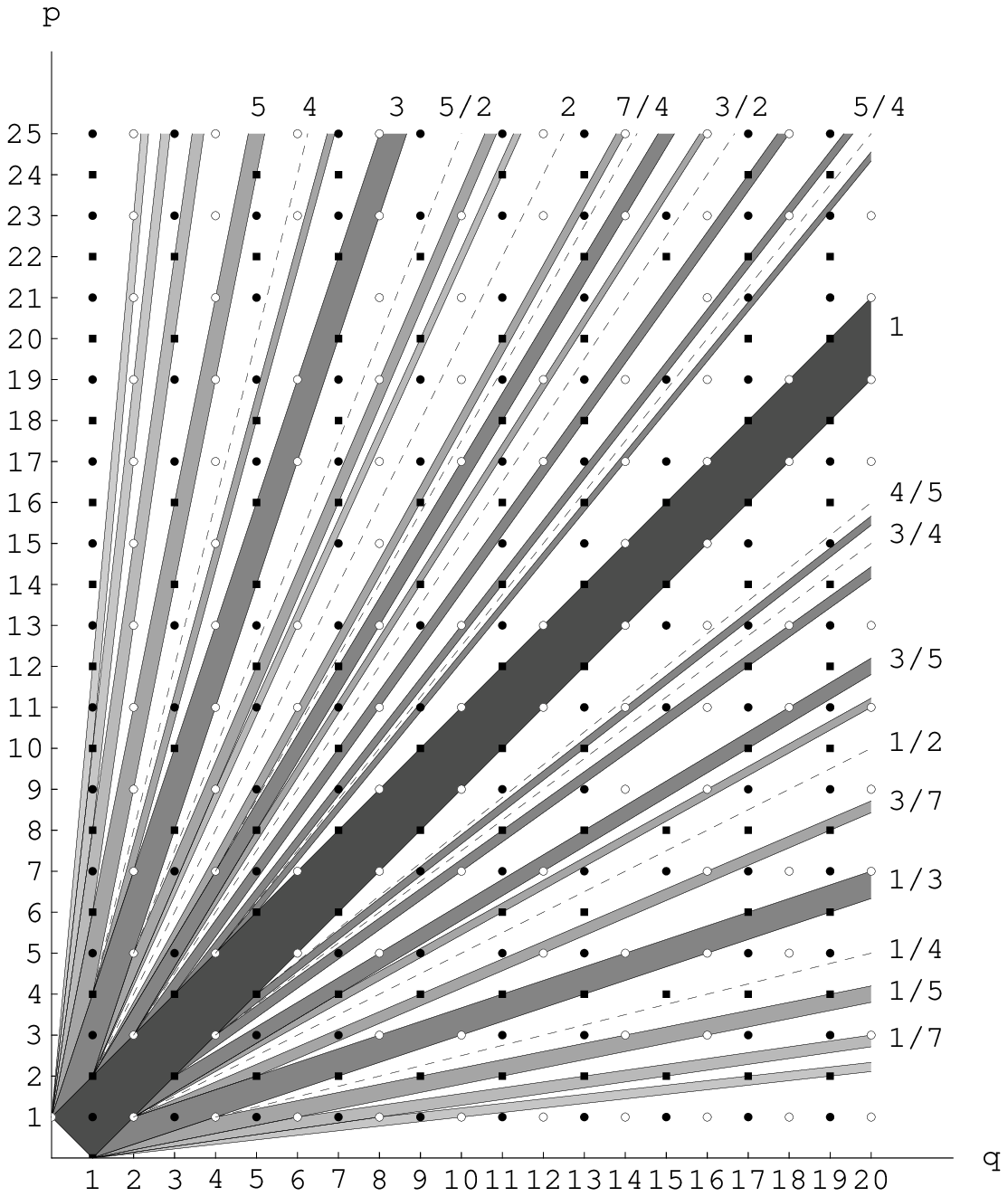}
}
\subfigure[The odd-even visibility diagram.]{
\label{fig2}
\includegraphics[width=0.47 \textwidth]{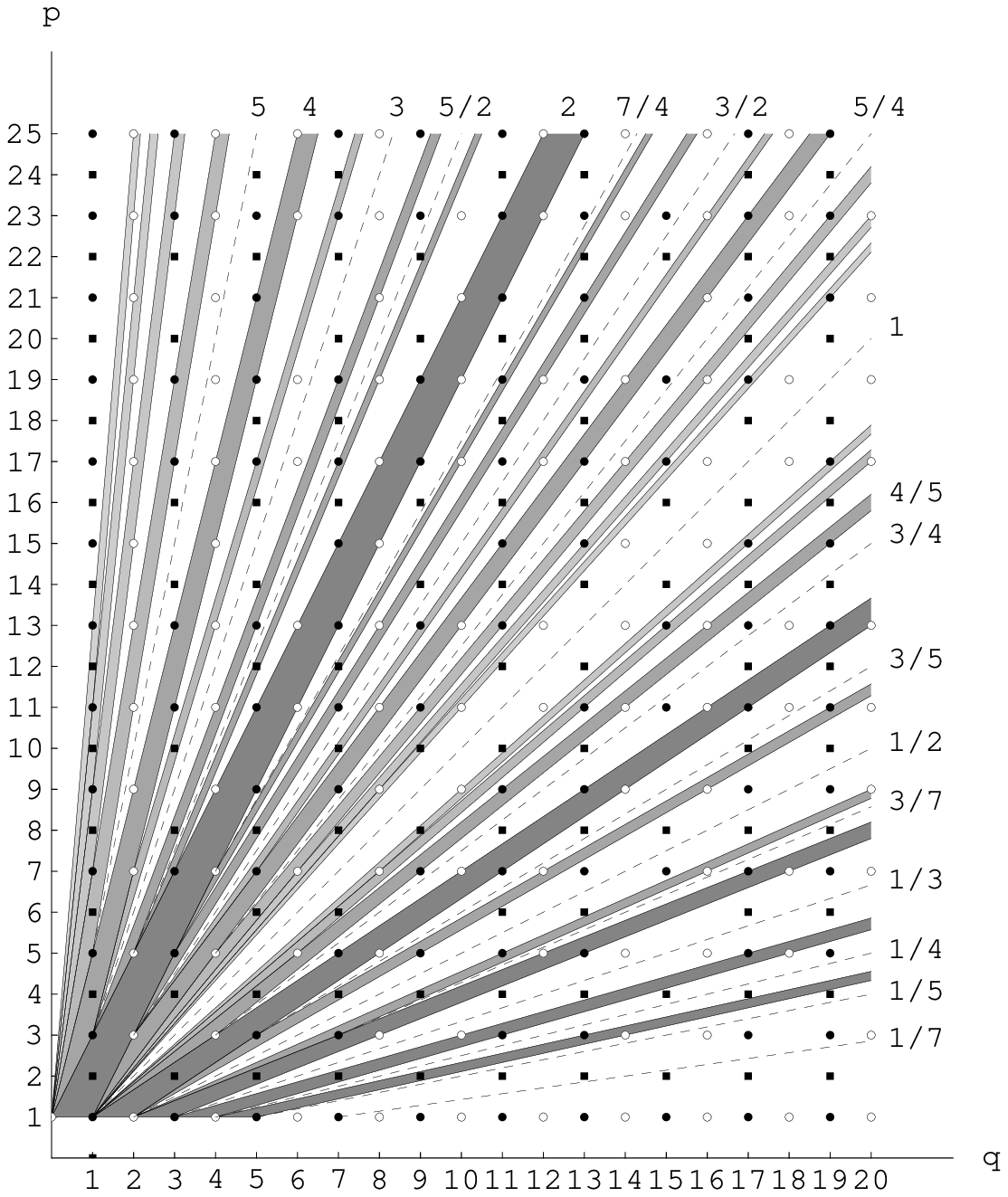}
}
\caption{The odd-odd and the odd-even visibility diagrams as taken from \cite{GMW3}. The parity of $q$ and $p$ are figured out by different symbols}
\end{figure}

\begin{figure}[ht]
\centering
\subfigure[The even-odd visibility diagrams.]{
\label{fig3}
\includegraphics[width=0.47 \textwidth]{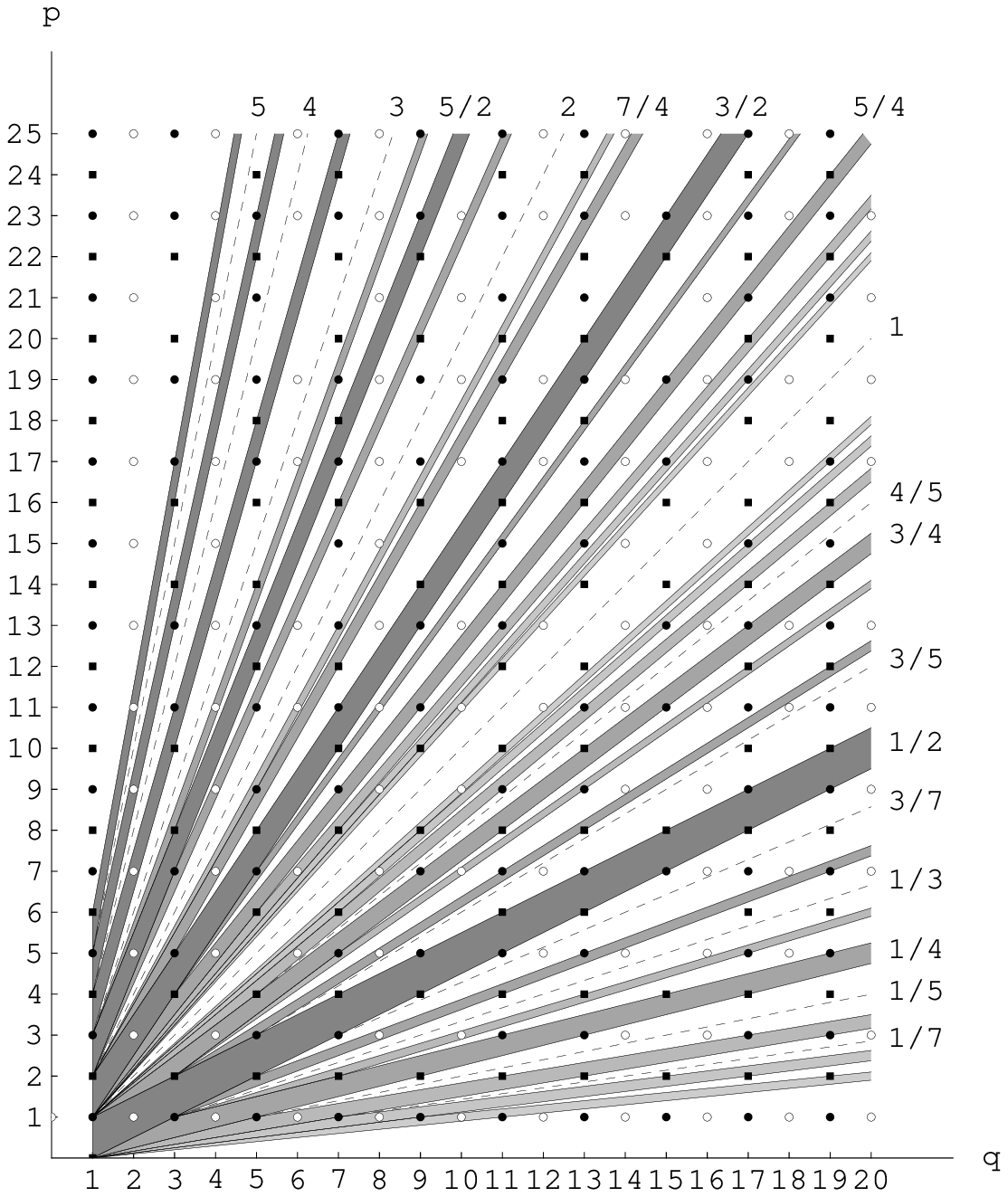}
}
\subfigure[The domains corresponding to the stable Hall phases. Notice the two white regions along the $p$ and $q$ axis which correspond respectively to $\nu=1/0$ (vertical ``stripe'') and to $\nu =0/1$ (horizontal ``stripe'').]{
\label{fig4}
\includegraphics[width=0.47 \textwidth]{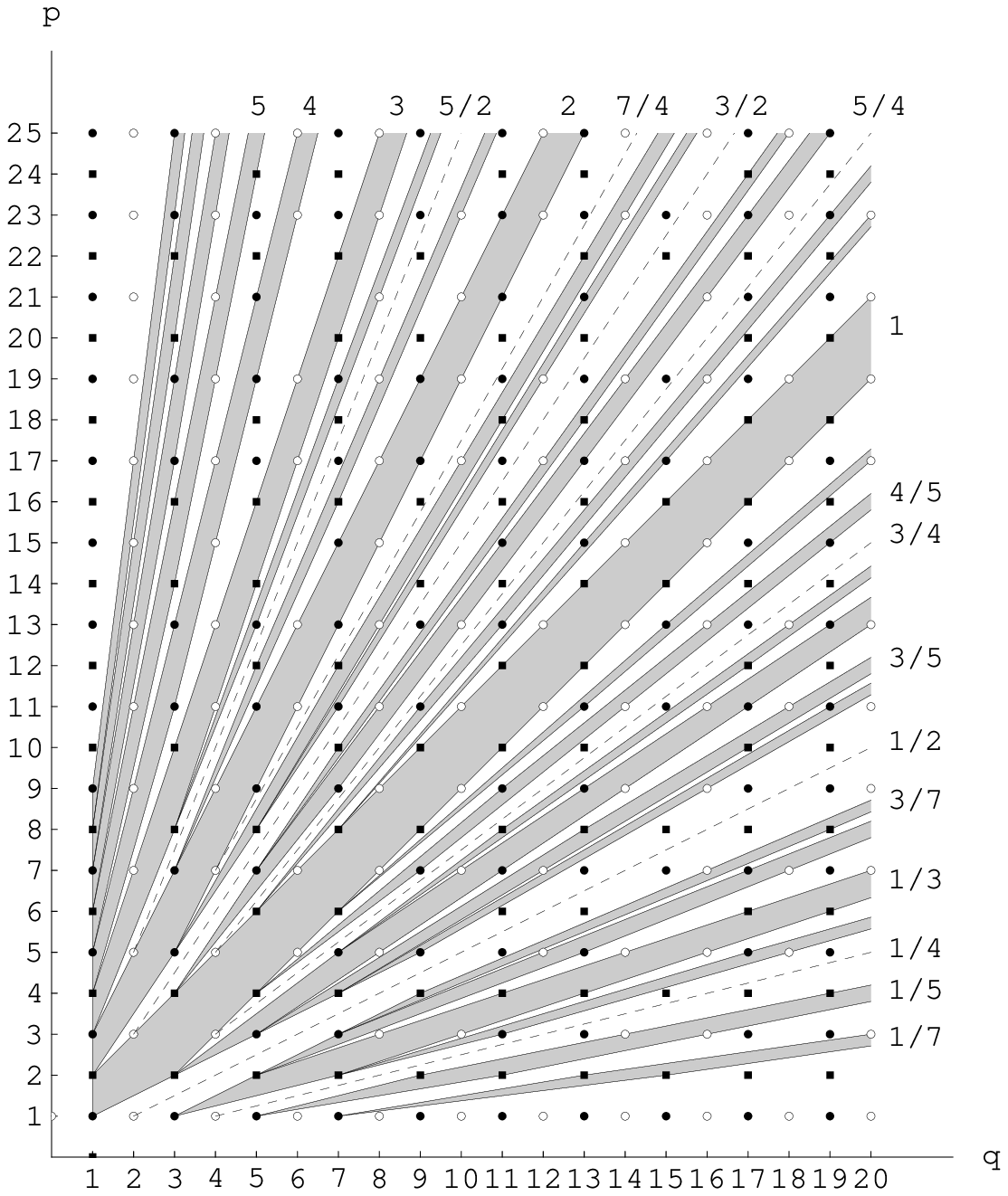}
}
\caption{Each gray stripe on the left diagram is in correspondance with a white zone in the right diagram.}
\end{figure}

The construction of the stripes in the visibility diagrams goes as follow. Consider the lattice of all the pairs of relatively prime positive integers $(q,p)$. It is convenient to distinguish the different pairs $(q,p)$ according to the even and/or odd character of its elements (hereafter called ``parity configuration''), namely $q$ odd and $p$ odd, $q$ odd and $p$ even, or $q$ even and $p$ odd. Then, consider the first parity configuration of these pairs ($q$ odd and $p$ odd). For each of them, one can use the above mentionned theorem to introduce all the pairs $(a,b)$ such that $pa - qb = 1$, and all the $(a',b')$ such that $pa' - qb' = -1$. It is very easy to show that the parity configuration of the pairs $(a,b)$ and $(a',b')$ are not odd-odd. These pairs can be displayed on the lattice, and are aligned on two half straight lines in this lattice. These two half straight lines define a stripe, which contains only the pair $(q,p)$. One obtains the ``odd-odd visibility diagram'' (figure~\ref{fig1}) by drawing the stripes associated to all the odd-odd pairs $(q,p)$ of the lattice. With the same procedure, it is possible to construct the ``odd-even visibility diagram'' (figure~\ref{fig2}) and the ``even-odd visibility diagram'' (figure~\ref{fig3}) if one considers exclusively the odd-even or the even-odd pairs $(q,p)$ of the lattice. Some usefull mathematical properties of these diagrams were presented in \cite{GMW3}. Let us only recall here that the structure of these 3 diagrams is connected to the modular group $\Gamma(2)$ (which, by the way, was our first motivation to construct these diagrams\footnote{The hypothesis of the group $\Gamma(2)$ as a possible group of symmetry for the QHE was investigated in \cite{GMW2}. There, it was showed that, in the holomorphic hypothesis, the renormalization group flows which preserve $\Gamma(2)$, reproduce a lot of the feature of the transitions in the QHE, and in particular the observed cross-over form. Moreover, a comparison with some similar results \cite{D} for the $\Gamma_0(2)$ modular group were performed.}), and that the (various proposed) hierarchies of the states of the QHE are encoded in the hierarchical structure of the stripes. In these visibility diagrams, each stripe is completely caracterized by a pair $(q,p)$, and we will index a stripe by this pair, or equivalently by its slope $p/q$. Obviously, the pair $(q,p)$ has the parity configuration of the visibility diagram considered.

In \cite{GMW1} and \cite{GMW3}, we were able to confront these diagrams to some experimental features of the QHE, and to define a mapping between them and physical quantities. We refer to these papers for more details. What has been learned there, is that the quotient $p/q$ is nothing but the filling factor ($\nu = N_c/N_\phi$, where $N_c$ is the number of charge carriers, and $N_\phi$ is the number of magnetic fluxes). More importantly, the $q$ and $p$ axis of the diagrams can be mapped respectively to the magnetic field $B$ (proportionnal to the number of magnetic flux $N_\phi$), and to the gate voltage $V_g$ of the Hall experiment. Recall that the gate voltage is used to control the number of charge carriers ($N_c$) in the Hall sample and that it can be qualitively related to the disorder by the simple relation: disorder $\sim 1/N_c$ (see for example in \cite{PG}). The $q-p$ plane involved in the diagrams can then be viewed as the $B-V_g$ plane of the QHE. The plateaus (the quantum Hall liquid states) of the QHE in the conductivity versus magnetic field plot were identified in this mapping with the stripes for which the filling factor is an odd denominator fraction. Notice that if one consider all these stripes in a single diagram (superposing the odd-odd and the odd-even diagrams), the stripes necesserally overlap. In the following section, we will use a general procedure to suppress the overlapping regions which is motivated by (and is consistant with) some experimental properties of the QHE. This will permit us to obtain a global phase diagram for the QHE.

\section{The construction of the global phase diagram}

The overlapping problem of the odd denominator stripes will be addressed using simple rules, some of which are based on experimental features of the QHE, some on minimal requirements. In particualr, these rules will give us the procedure to restrict the odd denominator stripes of the visibility diagram, in a consistent and general way to define the domain of the corresponding phase of the QHE.

First, recall that in any cross-over experiment, where the system goes from a (stable) quantum Hall state (plateau) to another one (transition in which a cross-over occurs between quantum Hall liquid states), there is always an extended region (even when the temperature goes to zero) where the system is neither in the first state nor in the second one (see for exemple \cite{HSSTX} and references therein). This experimental observation suggests that the phases for the stable states (the plateaus) cannot have a common line in their respective boundaries, which would correspond to a sharp transition from a plateau to another one. This last statement defines our first rule. Notice that in the so far experimental and theoretical approaches of the global phase diagram of the QHE, the separation lines between the quantum Hall liquids were essentially fixed arbitrarely, in general by choosing the cross-over point of the transition as the frontier between two states. Assuming this, the "states" between the quantum Hall liquid states are ignored, and the quantum Hall liquids are then adjacent. Here, the first rule permits one to take into account these intermediate phases.

Next, if one keeps in mind the above recalled physical identification of the axis of the visibility diagrams, at fixed disorder (\textsl{i.e.} for a given value $p_0$ of $p$), when the magnetic field is increased (\textsl{i.e.} $q$ is increased), one moves on these diagrams on a horizontal line $p=p_0$. For this value of the disorder, all the plateaus are not experimentally observed. Indeed, in our present scheme, this corresponds to the fact that these non observable plateaus, and therefore their (odd denominator) stripes, are associated to values of $\nu = p_1/q_1$ with $p_1>p_0$. They would become observable only for a lower value of the disorder. Consequently, the domain corresponding to a stable phase $\nu = p_1/q_1$ of the QHE must be in the upper half plane $p\geq p_1$, and is not experimentally observable if the actual value of the disorder $1/p_0$ is greater than $1/p_1$.  Now, a very similar analysis can be used to restrict this domain of a stable phase $\nu = p_1/q_1$ to the right half plane $q\geq q_1$ (if one consider the magnetic field in place of the disorder in the previous discussion). This defines our second rule.

Finally, our third rule is a minimal requirement about the alterations to perform on each odd denominator stripes in order to satisfy the first and second rules. Indeed, one can shrink in many ways each of these stripes to achieve this goal. We will only propose the simplest and minimal one. In particular, the same modification will be applied to all the odd denominator stripes without exception.

The second rule suggests that the $(q,p)$ vertex indexing an odd denominator stripe, is the lower left wedge of the phase domain. Two odd denominator stripes can only overlap near their indexing vertices. Indeed, they separate when $q$ and $p$ goes to infinity. So the first rule applies only in the vicinity of these vertices. There, a very simple definition of the domain can be given to obey this first rule. The resulting domains are depicted on figure~\ref{fig4}. The procedure we used there, which is the simplest one according to the third rule, is the following: the new boundary of the domain of the odd denominator state joins the indexing vertex $(q,p)$ of an odd denominator stripe to the nearest point $(a,b)$ on the boundary of the stripe with $a$ odd, $a\geq q$ and $b\geq p$. Notice that imposing $a$ to be odd has the effect to suppress some common lines in the boundaries of adjacent stripes. This is necessary according to the first rule.

\begin{figure}
\includegraphics[width=\textwidth]{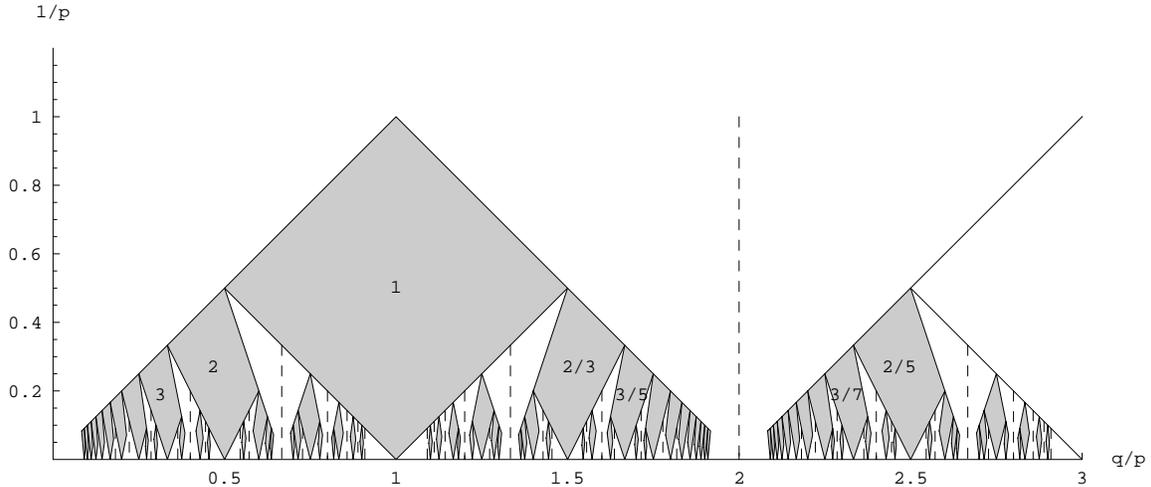}
\caption{The global phase diagram, in the variables \textsl{disorder} (identified with $1/p$) versus \textsl{inverse filling factor} (identified with $q/p$), obtained from the construction exposed in the text. Some phases are labelled by their corresponding values of $\nu$.}
\label{fig5}
\end{figure}

Figure~\ref{fig4} represents the proposed domains of the stable states (odd denominator filling fraction) of the QHE in the $q-p$ plane\footnote{Notice that this diagram has the same symmetries than the diagram obtained by superposing the odd-odd and odd-even visibility diagrams. So the conclusions of \cite{GMW3} about the possible groups of symmetrie for the QHE are not questionned by these modifications.}. The construction of the global phase diagram of the QHE is based on this last figure and on the previous mapping between the visibility diagram and the physical quantities relevant to the QHE. Obviously, without a microscopic model, finer details of this global phase diagram, for instance the exact form of the boundaries of the phases, cannot be obtained from this approach. What we obtain here is a global architecture of the phases, in particular their relative positions, and the topology of their boundaries. The global phase diagram of the QHE is generally drawn in \textsl{the disorder} (identified with the $1/p$ variable) versus \textsl{the inverse filling factor} (identified with the variable $q/p$) plane. The global phase diagram is then obtained from figure~\ref{fig4} using the mapping $(q,p) \mapsto (q/p, 1/p)$. The result is presented in figure~\ref{fig5}, where of course only a finite number of phases is depicted. Some dash lines of figure~\ref{fig4} (even denominator metallic states) are also reported on this diagram.

\section{Comments on the global phase diagram}

We would like now to comment the global phase diagram that we obtained. 

From the very construction, and the general properties of the visibility diagrams, the left zone of the diagram is an insulator phase (corresponding to $\nu\sim 1/0$). The upper region $1/p > 1$ can also be identified with an insulator. It has been suggested in the litterature \cite{SW} that these two insulator states could be different phases of the QHE. If these two insulator states really exist, they should be associated somehow with the two stripes indexed by $\nu=1/0$ (vertical stripe) and $\nu=0/1$ (horizontal stripe). These two stripes have not been drawn on any figure presented here, but they are nevertheless apparent on figure~\ref{fig4} as white regions along the $p$ and $q$ axis. 

Another global phase diagram for the QHE was proposed by Kivelson, Lee and Zhang in \cite{KLZ}. Their theoretical construction was based on the law of corresponding states. The general structure of our phase diagram bears some similarities with their phase diagram, in particular the location of the quantum Hall liquid states are roughly the same. Their shaded regions (figure~1 in \cite{KLZ}) correspond to our ``white'' regions ; they are centered about the half denominator states (the quantum Hall metallic states). As noticed in the previous section (first rule), their quantum Hall liquid states are adjacent, which is not the case in our diagram. This is due to the definition they take for the frontiers between two quantum Hall liquid states. The main structural difference between our diagram and the KLZ one lies in the possible \textsl{direct} transition between integer quantum Hall states and insulator, and their counterparts in other regions of the diagram (phases $\nu=2, 3, 2/3, 3/9, 3/7, \dots$). These transitions are strictly forbidden in the KLZ diagram, which is essentially the floating property.  One other theoretical phase diagram has been suggested by Halperin, Lee and Read in \cite{HLR}. This diagram has some very general common features with the KLZ one, but has the noticeable difference that it allows the direct transitions of the type  \textsl{integer quantum Hall states} $\leftrightarrow$ \textsl{insulator}, which are also permitted in our phase diagram. Our figure~\ref{fig5} is actually very similar, in many respects, to the figure~4 of \cite{HLR}, in particular, they is no floating in either one.

Concerning the experimental situation, from an analysis of several observed direct transitions, the authors of \cite{KMFP} proposed a synthetic phase diagram. The transitions stemming from the topology of our phase diagram are in good agreement with this general experimental picture shown there (figure~4 in \cite{KMFP}), except for the possible oscillations of the integer to insulator boundaries which are discussed in their paper. The author of \cite{HSSTX}, presented a more recent experimental phase diagram for the \textsl{integer} QHE. The topology of the integer phase diagram which can be extracted from our figure~\ref{fig5} is the same as their one. Notice that in this paper, the possible oscillations of the integer to insulator boundaries seems to be smoother compared to those appearing in \cite{KMFP}. Recent numerical simulations based on the TBM Hamiltonian gave rise to an integer phase diagram \cite{SW, SWW} very similar to the experimental one obtained by Hilke \textsl{et al.} In all these proposed phase diagrams, direct transitions \textsl{integer quantum Hall states} $\leftrightarrow$ \textsl{insulator} do exist\footnote{In \cite{WXNJ}, some numerical simulations based on an edge channels model, confirm also this feature.}. Let us stress the fact that the experimental situation is not yet cleared up \cite{SRKCWL}. In particular, the exact shape of the frontiers between the different quantum Hall phases seems to be notably dependant on the nature of the sample (MOSFETs, p-GaAs/AlGaAs, or p-SiGe) and/or on possible spin effects.

So far, we have insisted on the quantum Hall liquid states (odd denominator fractions). Concerning the metallic states (even denominator fractions), they appear in our phase diagram as simple straight lines surrounded by a whole family of unresolved states (white triangles in figure~\ref{fig5}) which are met in a typical cross-over transition from a quantum Hall liquid state to an other.

As a final remark, let us notice that there are some trajectories through the phase diagram which met only the integer quantum Hall liquid phases. This is indeed the case for any straigth line through the origin with slope between $1/3$ and $1$ (figure~\ref{fig5}). These trajectories can be generated experimentally if one increase the disorder, by acting on the gate voltage, while keeping fixed the magnetic field. This was actually the situation of the original integer quantum Hall experiment \cite{KDP}.

\section{Conclusion}

In this letter we have proposed a derivation of the global phase diagram for the QHE, based on visibility diagrams that we introduced in \cite{GMW1}. This phase diagram is in very good agreement with the experimental properties so far collected on the phases of the QHE and on their transitions. It is very close to one of the theoretical phase diagram which has been proposed in the litterature \cite{HLR}. It remains to explain the origin of this construction, in particular why some so simple principles give rise to such an accurate diagram. This will probably be achieved using either some workable mesoscopic phenomenological model (showing some modular symmetries) or some microscopic approach.

\end{document}